%% file: ms.tex
\documentclass[12pt,preprint]{aastex}

\newcommand{\msun} {$M_{\odot}$}

\newcommand{\halpha} {H$\alpha$}

\newcommand{\Te} {T_{\rm eff}}
\newcommand{\logg} {\log g}
\newcommand{\nh} {\log\ ({\rm H/He})}

\begin{document}

\title{Model Atmosphere Analysis of the Weakly Magnetic DZ White Dwarf G165-7}

\author{
P. Dufour\altaffilmark{1}
P. Bergeron\altaffilmark{1}, 
G. D. Schmidt\altaffilmark{2}
James Liebert\altaffilmark{2},
H. C. Harris\altaffilmark{3},
G.R. Knapp\altaffilmark{4},
S.F. Anderson\altaffilmark{5}, and
D.P. Schneider\altaffilmark{6}
}

\altaffiltext{1}{D\'{e}partement de Physique, Universit\'{e} 
de Montr\'{e}al, C.P. 6128, Succ. Centre-Ville, Montr\'{e}al,
Qu\'{e}bec, Canada H3C 3J7; dufourpa@astro.umontreal.ca,
bergeron@astro.umontreal.ca}
\altaffiltext{2}{Steward Observatory, University of Arizona, 933 North Cherry Avenue, Tucson, AZ 85721; gschmidt@as.arizona.edu, liebert@as.arizona.edu}
\altaffiltext{3}{US Naval Observatory, P.O. Box
1149, Flagstaff, AZ 86002-1149; hch@nofs.navy.mil}
\altaffiltext{4}{Princeton Univ. Obs., Peyton Hall,Princeton, NJ 08544; gk@astro.princeton.edu}
\altaffiltext{5}{Department of Astronomy, University of Washington, Box 351580, Seattle, WA 98195; anderson@astro.washington.edu}
\altaffiltext{6}{Department of Astronomy and Astrophysics, Pennsylvania State University, 525 Davey Laboratory, University Park, PA 16802; dps@astro.psu.edu}

\begin{abstract}
A reanalysis of the strongly metal-blanketed DZ white dwarf G165-7 is
presented. An improved grid of model atmospheres and synthetic spectra
is used to analyze $BVRI$, $JHK$, and $ugriz$ photometric observations
as well as a high quality Sloan Digital Sky Survey spectrum covering
the energy distribution from 3600 to 9000 \AA.  The detection of
splitting in several lines of Ca, Na, and Fe, suggesting a magnetic
field of $B_s\sim650$~kG, is confirmed by spectropolarimetric
observations that reveal as much as $\pm 7.5$\% circular polarization
in many of the absorption lines, most notably Na, Mg, and Fe.  Our
combined photometric and spectroscopic fit yields $\Te = 6440$~K,
$\logg=7.99$, $\nh=-3.0$ and $\log\ ({\rm Ca/He})=-8.1$. The
other heavy elements have solar ratios with respect to calcium, with
the exception of Na and Cr that had to be reduced by a factor of two
and three, respectively.  A crude polarization model based upon the
observed local spectral flux gradient yields a longitudinal field of
165~kG, consistent with the mean surface field inferred from the
Zeeman splitting. The inclusion of this weak magnetic field in our
synthetic spectrum calculations, even in an approximate fashion, is
shown to improve our fit significantly.

\end{abstract}

\keywords{stars: abundances -- stars: atmospheres -- stars: magnetic field
-- white dwarfs -- stars: individual (G165-7)}

\section{INTRODUCTION}

White dwarf stars of the DZ spectral type are cool, helium-atmosphere
objects showing traces of heavy elements in their optical
spectra. They are recognized mainly by the Ca~\textsc{ii} H $\&$ K
doublet, while a few also show Ca~\textsc{i} $\lambda$4226,
Mg~\textsc{i} $\lambda3835$ or Fe~\textsc{i} $\lambda3730$ (see
\citealt{SionAtlas}, \citealt{WAtlas}, and \citealt{harris03} for typical spectra). 
Since the heavy elements present in the atmospheric regions sink below
the photosphere in a time scale much shorter than the white dwarf
cooling time \citep{paquette86}, the presence of these metals in DZ
stars must be supplied by an external source, the most obvious
mechanism being accretion from the interstellar medium, or, in some
instances, from a circumstellar dust disk surrounding the white dwarf
\citep{zuckerman87,kilic05,becklin05}. Accretion of comets has also 
been proposed as a possible origin of metals in the atmospheres of DZ
stars \citep{Alcock86}. A precise determination of the heavy
element abundances observed in the optical or UV spectra of DZ stars
can thus provide important clues about the accretion rate and chemical
composition of the interstellar or circumstellar medium.

Atmospheric abundance analyses of DZ stars are usually based on a
restricted number of absorption lines detected in the optical portion
of the spectrum \citep[see, e.g.][]{LW87}, although a more important
set of lines can be used when UV spectroscopy is available
\citep{Zeidler86,KW00,WKL02}. G165-7 (WD~1328+307, LHS 2745, $V=16.03$) 
is a particularly unusual DZ white dwarf that has practically no flux
in the ultraviolet because of the severe blanketing by a large number
of metallic lines, but it is also characterized by a very rich optical
spectrum showing hundreds of absorption lines from various heavy
elements such as Ca, Cr, Fe, Mg and Na. The first model atmosphere
analysis of this star was carried out by \citet{WL1980} who found
$\Te=7500\pm 300$ K, $\logg=8.0\pm0.3$, $\nh<-4.0$, and abundances for
the heavy elements of about $1/30$ of the solar value under the
assumption of scaled solar abundances.  Their analysis was based on
low and moderate (3$-$5 \AA) resolution spectroscopy in several
segments covering the 3350$-$6700 \AA\ region, as well as on the
multichannel spectrophotometry of \citet{greenstein76}.

\citet[][hereafter BLR]{blr01} also analyzed the broad-band energy distribution 
(optical $BVRI$ \& near-infrared $JHK$ photometry) with a grid of pure
helium model atmospheres and found $\Te=7320$~K and $\logg=8.21$ (the
surface gravity in this case is constrained by the trigonometric
parallax). However, analyses of the energy distribution of DQ white
dwarfs based on pure helium models tend to overestimate the effective
temperature by 300-800~K and the surface gravity by 0.10-0.24 dex with
respect to the atmospheric parameters obtained from models that
include carbon. This effect is understood in terms of an increase of
the He$^-$ free-free opacity resulting from the additional free
electrons \citep{provencal02,dufour05}. A similar effect is thus
expected in the case of DZ stars.

We present here a reappraisal of the analysis of G165-7 based on our
latest metal blanketed atmospheric models and a new high quality
spectrum. The observations are described in \S~\ref{observation} while
in \S~\ref{theoretical} we describe our theoretical framework
including our model atmosphere and synthetic spectrum
calculations. Our detailed analysis is presented in \S~\ref{analysis}
and our conclusions follow in \S~\ref{conclusion}.
 
\section{OBSERVATIONS}\label{observation}

Our analysis relies on both spectroscopic and photometric observations
from various sources. The optical spectrum secured by BLR
(details of the observations are also provided there) covers the
3500$-$8300 \AA\ region at a resolution of 6~\AA\ FWHM.  A
new spectrum has been obtained by the Sloan Digital Sky Survey
(SDSS), where the star is detected as SDSS J133059.26+302953.2.  
The SDSS \citep{york00,stoughton02} obtains five-band imaging
\citep{gunn98,lupton01,pier03,gunn06} from which objects are selected
for spectroscopic observations. The SDSS quasar survey targets point
sources with non-stellar colors \citep{richards02}, some of which,
like G165-7, turn out to be peculiar stars. The spectrum was observed
on 7 April 2005. It covers a wavelength range of
3800$-$9200 \AA\ at a resolution of $\sim 3$ \AA\ FWHM. The
comparison of the two spectra shown in Figure \ref{fg:f1} reveals that
the agreement in terms of {\it absolute} fluxes is surprisingly good,
apart from the red portion of the BLR spectrum, which suffers from an
obvious flux calibration problem. Also indicated at the top of the
figure are the most prominent detected absorption features.
Most interestingly, a closer examination of the SDSS spectrum
reveals the presence of splitting in several atomic lines
(e.g.~Fe~\textsc{i} $\lambda\lambda5269-5328$, Na~\textsc{i} D and
Ca~\textsc{ii} $\lambda\lambda8542-8662$), suggesting the presence of
a weak magnetic field (see below).

BLR also reported Cousins $BVRI$ and CIT $JHK$ photometry for
G165-7. Also available are SDSS photometric observations in the
$ugriz$ system \citep{fukugita96, hogg01,smith02,ivezic04}. The
photometric observations that will be used to determine the effective
temperature are reported in Table 1 together with the trigonometric
parallax measurement from the Yale parallax catalog \citep{ypc}. The
latter will be used to constrain the radius, and thus the surface
gravity (or the mass) of the white dwarf (see \S~\ref{analysis}).
Infrared 2MASS photometry is also available for G165-7:
$J=15.402\,(0.044)$, $H=15.282\,(0.087)$, and $K_S=15.413\,(0.135)$;
although these values are consistent with the BLR measurements
reported in Table 1, the corresponding 2MASS errors are larger for $H$ and
$K_S$ and they will not be considered further.

\section{THEORETICAL FRAMEWORK}\label{theoretical}

Our LTE model atmosphere code is similar to that described by
\citet{dufour05} for the study of DQ white dwarfs. It is based on a
modified version of the code described at length by
\citet{BSW95}, which is appropriate for pure hydrogen and pure
helium atmospheric compositions, as well as mixed hydrogen and helium
compositions, while energy transport by convection is treated within
the mixing-length theory. One important modification is that metals
and molecules are now included in our equation-of-state and opacity
calculations \citep[see][for details]{dufour05}. As was the case for
DQ white dwarfs, He$^-$ free-free is found to be the dominant source
of opacity in DZ stars. It is thus important to include all
possible sources of electrons in the equation-of-state, and we have
included here all elements with Z $\le$ 26. The chemical abundances
cannot be determined individually, however, since most of these
elements are not observed spectroscopically. We thus initially assume
that the relative abundances are consistent with solar ratios, a
reasonable assumption according to 
\citet[][see also Wolff et al. 2002 for other DZ stars]{WL1980} and our own preliminary analysis based on a chemical
abundance analysis of the observable elements.

Our model grid covers a range of atmospheric parameters from
$\Te=4500$ to 12000~K in steps of 500 K, and $\log\ ({\rm Ca/He})=
-7.0$ to $-12.0$ in step of 0.5 dex, while $\logg$ is kept fixed at
8.0, an assumption that will need to be verified a posteriori.  This
grid contains no hydrogen. The relative abundances of all
elements heavier than helium are set with respect to the calcium
abundance in solar ratios. Additional models described in
\S~\ref{analysis} have been calculated with different relative
abundances of metals, models including hydrogen, and in particular
models taking into account the presence of a weak magnetic field.

In the context of cool DZ stars, in addition to the increased He$^-$
free-free continuum opacity, important ultraviolet absorption features
may potentially affect the energy distribution, and thus the
atmospheric structure, with respect to pure helium models. Thus, over
4000 of the strongest metal lines --- $\sim 2600$ lines from
Fe~~\textsc{i} alone --- are included explicitly in both the model and
synthetic spectrum calculations. These lines were selected by
taking all lines contributing more than one tenth of the He$^-$
free-free opacity in the range $\tau_R = 0.1 - 1.0$ from several models
at $\log$ Ca/He=$-$7 and $\Te$ between 5000 K and 12,000 K. We are
confident that our line list includes all the important contributors
to the atomic opacity since spectra calculated by increasing the
number of lines by an order of magnitude did not have any effect on
the emerging spectrum. The line absorption coefficient is calculated
using a Voigt profile for every line and every depth point. The line
broadening is treated within the impact approximation with van der
Waals broadening by neutral helium. Central wavelengths of the
transitions, $gf$ values, energy levels, and damping constants are
extracted from the GFALL line list of R.~L.~Kurucz\footnote{see
http://kurucz.harvard.edu/LINELISTS.html}.  We will show in
\S~\ref{magnetic}, however,  that magnetic broadening is the dominant broadening mechanism in the
case of G165-7.

\section{DETAILED ANALYSIS}\label{analysis}

\subsection{Atmospheric Parameter Determination}

The first step in the analysis of the observational data is to
estimate the effective temperature of the star. Our fitting procedure is
similar to that used for our analysis of DQ white dwarfs
\citep{dufour05}. Briefly, we transform the magnitudes at each
bandpass into observed average fluxes $f_{\lambda}^m$ using the
following equation

$$m= -2.5\log f_{\lambda}^m + c_m\ , \eqno (1)$$

\noindent where the values of the constants $c_m$ are taken from
the latest photometric calibration of \citet{HB06} based on the
absolute fluxes of Vega observed by the {\it Hubble Space Telescope}
with the {\it Space Telescope Imaging Spectrograph}
\citep{bohlin04}. The resulting energy distributions
are related to the model fluxes through the relation

$$f_{\lambda}^m= 4\pi~(R/D)^2~H_{\lambda}^m\ , \eqno (2)$$

\noindent
where $R$ is the radius of the star, $D$ its distance from Earth, and
$H_{\lambda}^m$ is the Eddington flux --- which depends on $\Te$,
$\logg$ and the chemical abundances, properly averaged over the
corresponding filter bandpass. The fitting procedure relies on the
nonlinear least-squares method of Levenberg-Marquardt based on a
steepest decent method \citep{pressetal92}. The value of $\chi ^2$ is
taken as the sum over all bandpasses of the difference between the two
sides of equation (2), properly weighted by the corresponding
observational uncertainties.

We begin by fitting the $BVRI$ and $JHK$ photometry to obtain an initial
estimate of the effective temperature (the $ugriz$ photometry is not
used at this stage). We assume $\logg=8.0$ throughout. Since the
temperature obtained from the energy distribution depends on the
assumed chemical composition, we rely on the spectroscopic
observations to constrain the metal abundances. We thus assume the
$\Te$ value obtained from the energy distribution and determine the
chemical composition by fitting the SDSS spectrum with our grid of
synthetic spectra, properly convolved with a Gaussian instrumental
profile at 3~\AA\ FWHM. The fitting procedure relies on the
least-squares method of Levenberg-Marquardt. A new estimate of the
effective temperature is then obtained by fitting the photometric
observations with models interpolated at the metal abundances
determined from the spectroscopic fit (all metal abundances are solar
with respect to calcium). The procedure is then iterated until the
atmospheric parameters have converged to a consistent photometric and
spectroscopic solution.

\subsection{Adopted Atmospheric Parameters}

As discussed below, the difficulties encountered in fitting the blue
region of the optical spectrum caused us to discard for the time
being the $B$ magnitude
from our fitting procedure (for the same reason, we fitted
only the 5000$-$9000 \AA\ region of the SDSS spectrum). The long
baseline provided by the optical $VRI$ and near-infrared $JHK$
photometry yields a sufficiently reliable estimate of the effective
temperature.  The photometric and spectroscopic fitting technique
described above yields $\Te=6440\pm 210$ K and $\log\ ({\rm Ca/He})=
-8.1\pm0.15$ (with all heavy elements assumed to be solar with respect
to calcium), where the uncertainties are obtained from the covariance
matrix of the fit; our photometric fit will be shown later
(\S~\ref{magnetic}). The stellar radius is derived from the
solid angle in equation (2) combined with the distance $D$ obtained
from the trigonometric parallax measurement. The radius is then
converted into $\logg$ (or mass) using evolutionary models similar to
those described by
\citet{fon01} but with C/O cores, $q({\rm He})\equiv
\log M_{\rm He}/M_{\star}=10^{-2}$, and $q({\rm H})=10^{-10}$, 
which are representative of helium-atmosphere white dwarfs.  We find a
value of $\logg=7.99\pm0.29$, entirely consistent with our initial
assumption of $\logg=8$ for our model grid. This corresponds to a mass
of $0.57\pm0.17$ \msun, in agreement with the mean mass of white dwarf stars
\citep{BSL,liebert05}. Our value of the effective temperature is significantly smaller
than the value of $\Te=7320$~K obtained by BLR under the assumption of
pure helium-atmosphere models. This is a direct consequence of the
increased He$^-$ free-free opacity in our model calculations. Because
of their higher temperature estimate, BLR required a smaller
solid angle to match the photometric obervations, which translated
into a smaller radius and thus a higher surface gravity of
$\logg=8.21$ (or 0.71 \msun).

In the remainder of this analysis, we adopt the atmospheric parameters
determined above. As mentioned in the previous section, the abundances
of all heavy elements are assumed to be solar with respect to calcium,
even if they are not detected in the spectrum.  In order to test the
sensitivity of this assumption on our atmospheric parameter
determination, we also calculated a model with our adopted parameters,
but by setting the abundances of all non-detected elements to
zero. The resulting spectrum is almost identical to that obtained
under the assumption of solar composition with respect to calcium.
This is perhaps not too surprising since the non-detected elements do
not contribute significantly to the electron population (see
below). Hence we feel confident that our basic assumption does not
affect the atmospheric parameter determination.  In what follows, we
will also allow some changes in the abundances of elements observed
spectroscopically (in particular hydrogen). But at the end, we will
again measure the impact of these changes on our atmospheric parameter
determination.

Although the assumption of metallic abundances scaled with respect to
solar abundances is appropriate for fitting globally the 5000$-$9000
\AA\ region, individual lines such as the Na $\lambda 5892$ resonance
doublet, the Cr $\lambda5208$ and the Fe ($\sim 5200-5500$~\AA)
absorption features are predicted to be stronger than observed, and
the abundances of these elements must be reduced by factors of 2, 3,
and 3, respectively, to produce an acceptable fit. Our best resulting
fit to the SDSS spectrum is shown in Figure \ref{fg:f2} and at a
higher resolution in Figure
\ref{fg:f3}. The overall spectrum in Figure \ref{fg:f2} is generally 
well reproduced by our model. In particular, the
predicted continuum slope in the red part of the spectrum is entirely
consistent with our effective temperature determination, which is
based on $VRI$ and $JHK$ photometric observations. This provides an
internal consistency check between our temperature determination and
the photometric and spectroscopic observations.

A more detailed comparison of our spectroscopic fit reveals various
discrepancies, however. Most noteworthy, the predicted flux in the blue region
of the spectrum between 3600 and 4600 \AA\ is considerably larger
than that observed.  This region can be better reproduced either by
reducing the effective temperature or by increasing the iron
abundance, but in both cases, the region longward of 4500 \AA\ cannot
be matched properly, including the slope of the energy distribution as
well as the iron absorption lines between 5200 and 5500 \AA, which are
then predicted to be much stronger than observed. The UV flux
deficiency could perhaps be explained by a missing source of opacity
that would partially mask the absorption lines in this region. We have
explored adding more lines, continuous opacities from heavy elements,
the occupation probability formalism of
\citet{HM88}, as well as the Ly$\alpha$ wing calculations of
\citet{KW00}, but none of these mechanisms were able to reduce the UV flux
significantly.

Figure \ref{fg:f3} also reveals that most lines are predicted to be
narrower and deeper than observed. Our model even predicts some sharp
Ca~\textsc{i} features around 6160 and 6450 \AA\ that are not observed
spectroscopically. The Mg~\textsc{i} $\lambda5175$ triplet is also not
well reproduced. This asymmetric profile has been interpreted by
\citet{WL1980} as evidence for quasi-static van der Waals
broadening; we postpone our discussion of this feature to
\S~\ref{mgh}. Moreover, there appears to be some missing line contribution
in the region around 4380 and 4400 \AA.  Even the location
of some absorption lines seems out of place.  For instance, a careful
examination of the region near 4050, 4150 and 4260
\AA\ indicates that the strongest predicted features are not matched
exactly. These small line shifts cannot be accounted for by
uncertainties in the SDSS wavelength calibration since the BLR and
SDSS spectra agree so perfectly (see Fig.~\ref{fg:f1}). The observed
discrepancies could not be improved by including additional metallic
lines in our calculations; therefore another explanation must be
sought.

\subsection{The Presence of a Weak Magnetic Field}\label{magnetic}

The SDSS spectrum in Figure \ref{fg:f1} suggests
the presence of Zeeman splitting, and a direct comparison of the
observed spectrum with our non-magnetic synthetic spectrum in Figure
\ref{fg:f3} clearly reveals line splitting. The most obvious cases
are the Fe~\textsc{i} $\lambda\lambda5269-5328$, Na~\textsc{i} D, and
the Ca~\textsc{ii} $\lambda\lambda8542-8662$ absorption lines. The
Fe~\textsc{i} lines near 5400 \AA\ are also not well reproduced by our
model spectrum. The peculiarity of the iron features observed here is
striking when compared with the spectrum of NLTT 40607 \citep[see
their Fig. 4]{Kawka04}, another rare cool DZ star similar to G165-7
which, however, does not show any evidence of line splitting.

To our knowledge, the only polarimetric test for a magnetic field on
G165-7 is the broadband ($3100-8600$ \AA) circular polarimetry
obtained by \citet{angel81} that yielded $V/I=+0.071\pm0.075$\%, or
an estimated mean longitudinal field strength $B_e = 230\pm240$~kG.
New polarimetric measurements were therefore obtained on 2005
December 30 UT and 2006 May 3 UT using the Steward Observatory 2.3m
telescope and SPOL spectropolarimeter of GDS.  The instrument was
configured for low spectral resolution ($\Delta\lambda=17$ \AA) but
broad coverage ($4200-8200$ \AA); calibration and reduction of the
600~s observational sequences were carried out in the usual manner
\citep[e.g.,][]{Schmidt92b}.  The results for 2005 Dec., shown in
Figure \ref{fg:f4}, conclusively verify the presence of a magnetic
field, with sharp undulations in circular polarization at each of the
prominent lines, reaching $V/I=\pm7.5\%$ at the strongest Mg, Na, and
Fe features.  Shown by the bold line in Figure \ref{fg:f4} is a
synthetic polarization spectrum computed for $B_e=165$~kG from a
simple model based on the local spectral flux gradient
\citep[e.g.][]{angel81}.  The modeling technique is adequate for the
linear Zeeman effect in the wings of lines but fails in the resolved
cores, and thus tends to underestimate the peak excursions in
polarization.  For a centered dipole field configuration with no limb
darkening, the ratio of the mean longitudinal to mean surface field
strength $B_e/B_s$ ranges between $\pm0.25$ for the full range of
inclination angles to the dipole axis, therefore $B_e$ is a sensitive
indicator of orientation.  Analysis of the individual
spectropolarimetric observations yields values of
$B_e=162,~140,~156$~kG, with an uncertainty of $\sim$15~kG for each
measurement.  Thus, there is no evidence for rotation over the 30 min
period covered by the observations.  However, a clear change occurred
between the two epochs, as the amplitude of the polarimetric features
in 2006 May is reduced by more than a factor two.  For comparison
with the \citet{angel81} measurement, the net circular polarization
of the data in Figure \ref{fg:f4} is $-0.034\pm0.030\%$, i.e., not
significantly different from zero.

In general, detailed calculations of synthetic spectra in the
presence of a magnetic field require a special numerical and physical
treatment that cannot be easily implemented in a non-magnetic model
atmosphere code. However, if we assume that the magnetic field does
not affect the atmospheric structure significantly, some simple
approximations can be used to include its effects on the emergent
spectrum only. For a relatively weak field, which is certainly the case
here, we can use the approximation of the linear Zeeman effect to
estimate the splitting of the atomic lines. In this regime, the
wavelength separation (in \AA) between the
centroids of the $\sigma$ components is given by the simple formula

$$\Delta\lambda= 9.34\times10^{-13} \lambda_c^2 g_{\rm eff} B_s\ ,\eqno (3)$$

\medskip
\noindent where $\lambda_c$ is the central wavelength of the 
line and $g_{\rm eff}$ is the effective Land\'e factor. We estimate
the mean surface field strength from an average value of the line
splitting measured for  Fe~\textsc{i} $\lambda\lambda$5269, 5328, 4383, and Na
$\lambda5892$, and by assuming $g_{\rm eff}=1.0$ for all lines.  With this
very simple approximation, we find a mean surface magnetic field of
$B_s\sim650$~kG, a value consistent with
the mean longitudinal field of $B_e=165$~kG estimated above.  Fields
of this strength are not unusual among white dwarfs, but G165-7
represents a rare case of a DZ white dwarf that exhibits sharp enough
metallic features to make the magnetic field detectable through Zeeman
splitting. Only two other DZ stars are known to exhibit such metallic
line splitting: LHS 2534 with $B_s = 1.92$ MG \citep{Reid01} and SDSS
J015748.15$+$003315.1 at $B_s = 3.7$ MG \citep{Schmidt03}.

In order to take into account, even approximately, the presence of the
magnetic field in our model calculations, we simply assume that all
spectral lines are split into three components of equal strengths
separated by a value of $\Delta\lambda$ given by equation (3), with
each component having one third of the total line strength in the
zero-field calculation \citep[see also][]{Schmidt92a}. We then
recalculate a synthetic spectrum at our previously determined
atmospheric parameters by using this approximate treatment.

The resulting spectroscopic fit using the same
chemical abundances as above indicates that the iron abundance no
longer needs to be reduced by a factor of 3, and that a solar
abundance of iron relative to calcium now matches the 5200$-$5500 \AA\
lines quite well, as shown in Figure \ref{fg:f5}. The improvement over
our non-magnetic calculations can be appreciated by contrasting these
results with those shown in Figure \ref{fg:f3}.  One important
consequence of the increased iron abundance is that the flux
deficiency problem in the blue region of the spectrum (top panels of
Figs.~\ref{fg:f3} and \ref{fg:f5}) has been dramatically improved,
although it is admittedly not perfect yet.  The missing line contribution in
the 4380$-$4400 \AA\ region discussed above is also naturally
explained. There are still a few residual discrepancies, however,
particularly near the 4100 and 4350 \AA\ regions, but given the crude
approximation we used here to take into account the presence of the
magnetic field, we feel our solution is quite satisfactory.

The other panels also reveal a better match for several metallic lines,
in particular Na~\textsc{i} D and Ca~\textsc{ii} $\lambda\lambda8542-8662$.  The
inclusion of a weak magnetic field also solves the problems of
the sharp Ca~\textsc{i} absorption features around 6160
and 6450 \AA\ (third panel), since the line splitting has now
broadened these features considerably. There is a small discrepancy
near 6560~\AA\ that is most naturally explained by a contribution from
hydrogen (H$\alpha$; see next section).

Finally, we show in Figure \ref{fg:f6} our best fit to the $(B)VRI$
and $JHK$ photometric observations using model fluxes that take into
account the presence of a 650 kG magnetic field. Also displayed are
the $ugriz$ photometric observations converted into average fluxes
using the transformation equations provided by
\citet{HB06}. The complete energy distribution is well reproduced with
our model, with the exception of the $u$ band, which may suffer from
atmospheric extinction problems \citep[see][]{HB06}.

\subsection{Hydrogen Abundance Determination}

\citet{WL1980} place a limit on the hydrogen abundance of $\nh<-4$
in G165-7 based on the absence ($W<2$~\AA) of an H$\alpha$ absorption
feature. This limit is very sensitive to the assumed
effective temperature of the star. Our effective temperature
determination is $\sim 1000$~K cooler than that obtained by
\citet[][$\Te=7500$~K]{WL1980} so the limit on the hydrogen abundance
needs to be increased.  Furthermore, a close inspection of the SDSS
spectrum, together with our spectroscopic fit (Fig.~\ref{fg:f5}),
clearly reveals the presence of a very shallow and broad H$\alpha$
absorption feature. This observed width is actually what is expected at these
temperatures since van der Waals broadening by neutral helium is the
main broadening mechanism in the physical conditions encountered
here. Note also that without a good modeling of the calcium lines in
the blue wing of H$\alpha$, this absorption feature could easily be
interpreted as noise. The H$\alpha$ feature cannot be detected in the
lower signal-to-noise ratio spectrum displayed in Figure 3 of
\citet{WL1980}.

We find that a model spectrum with $\nh=-3$ reproduces almost
perfectly the H$\alpha$ region (see the insert in Fig.~\ref{fg:f5}).
Here, the magnetic field for H$\alpha$ has been included following the
procedure outlined in \citet{bergeron92}, although its effect 
on the observed profile is found to be negligible since
van der Waals broadening by neutral helium completely dominates in
this physical regime.  Because of the high metallicity content
determined for G165$-$7, the contribution of hydrogen to the electron
population is relatively small, except perhaps in the deepest layers
where the temperature is high enough to produce substantial
ionization. Actually, at the photosphere ($\tau_R\sim 1$), the
principal electron donors are Mg (27.6\%), Fe (21.5\%), Si (17.1\%)
and H (24.8\%), while Ca and Na contribute 2.0\% and 1.7\%,
respectively. Furthermore, the contributions of the H$^-$ opacity and
H$_2$ collision induced opacity are negligible compared to the
dominant He$^-$ free-free opacity. In order to estimate how our
atmospheric parameter determination would change given (1) the
presence of hydrogen, (2) the reduced abundances of Na and Cr, and (3)
the presence of the magnetic field, we fitted our best model spectrum
shown in Figure \ref{fg:f5} with our hydrogen-free, non-magnetic,
solar scaled models. The atmospheric parameters we obtain differ by
only 100~K in $\Te$ and 0.2 dex in $\log\ ({\rm Ca/He})$.

Hydrogen, being the lightest element, tends to raise to the upper
layers of the photospheric regions of the star, and its abundance
should thus always {\it increase} with time as the white dwarf experiences
multiple episodes of accretion from the interstellar
medium. Previous studies
\citep[e.g.,][]{dupuis3,WKL02} have shown, however, that DZ stars have less 
hydrogen in their atmospheres than what is expected if the accreted
material has a solar composition. Consequently, the hydrogen accretion
rate must be reduced relative to that of metals in order to
account for the relatively low hydrogen abundances (with respect to
heavier elements) observed in DZ stars. One model,
proposed by \citet{WT82} to reduce the accretion of hydrogen, is the
so-called propeller mechanism. In this model, protons are prevented
from accreting onto the surface of the white dwarf by a rotating
magnetic field, while metals, most probably in the form of grains, are
unaffected by this mechanism and thus reach the surface. In addition,
this model requires a relatively strong UV radiation flux to ionize
hydrogen in the immediate surroundings of the star, a requirement that
is certainly not met in the case of G165-7.  Since the propeller
mechanism cannot prevent hydrogen from being accreted onto the surface
of G165-7, we expect the metal-to-hydrogen ratio to be much lower than
the solar value, $\log\ ({\rm Ca/H})_\odot=-5.6$.  However, the value
we determined for G165-7, $\log\ ({\rm Ca/H})=-5.1$, is slightly
larger than the solar value.  G165-7 thus represents another case
of a DZ star with a unexpectedly low hydrogen abundance.

\subsection{Asymmetric MgI Line or MgH Molecular Absorption?}\label{mgh}

Based on the absence of H$\alpha$ and their relatively high effective
temperature estimate, \citet{WL1980} rejected the possibility that the
asymmetric Mg~\textsc{i} $\lambda$5175 ``b'' profile could be
explained by a MgH molecular absorption band. Instead, they
interpreted the observed asymmetry as evidence that the impact theory
was not valid anymore for this line. A quasi-static profile was found
to give a reasonable agreement to the shape of the blue wing of the
Mg~\textsc{i} line if the C$_6$ damping constant was increased by a
factor 120. The lower temperature and higher hydrogen abundance
determined here allow us to reconsider this interpretation.

In order to explore this alternative interpretation, we have
explicitly included the MgH line opacity in our calculations. Although
our code takes into account the formation of molecules in the
equation-of-state, the molecular opacity for the coolest stars has not
been fully implemented yet, with the exception of the C$_2$ molecular
Swan bands required for analysis of DQ stars \citep{dufour05}. Our
result, shown in Figure
\ref{fg:f7}, indicates that it is indeed possible to match approximately 
the blue wing at $\sim 5100$ \AA, but only by increasing the MgH
abundance by a factor of $\sim 20$. We can only offer tentative
explanations to explain this inconsistency.

One possibility is that because of a missing ingredient in our model
calculations --- an important source of opacity or additional electron
donors for instance, our effective temperature determination is still
overestimated. If G165-7 is actually cooler than suggested by our
models, more hydrogen could be present and molecular formation would
be favored. For example, at $\Te=6250$~K, the above factor is reduced
to 10, but the overal energy distribution predicted is not reproduced
as well as with our adopted 6440~K solution. Alternatively, the
molecular absorption process could be affected by the magnetic field,
but detailed calculations are needed to confirm this hypothesis. We
cannot rule out the possibility that we are simply dealing with the
breakdown of the impact approximation, as initially proposed by
\citet{WL1980}, but until this feature is correctly fitted within a
coherent theoretical framework, we must conclude that the nature of
the Mg~\textsc{i} asymmetric profile remains uncertain. We note that
it is unlikely that non-ideal effects in the equation-of-state
(Coulomb corrections, finite volume effect, line quenching) are
important since the pressure in our model never exceeds $10^{10}$
dynes cm$^{-2}$.  Based on a modified version of the equation-of-state
of \citet{Mihalas90}, \citet{WKL02} found only a weak influence in van
Maanen 2, another DZ star in which the atmospheric pressure is
certainly higher than that in G165-7 since it is cooler and possesses a
smaller metal content. We note finally that the sample of DZ stars
displayed in Figures 10 and 11 of \citet{harris03} show several objects
with similar asymmetric profiles, some of them certainly too hot
($\sim8000$~K) to be explained by molecular absorption.

\section{SUMMARY AND CONCLUSIONS}\label{conclusion}

Our study of G165-7 represents the first detailed model atmosphere
analysis of a magnetic DZ white dwarf. High quality spectroscopic
observations revealed that several metallic features were split by a
surface magnetic field of approximately 650 kG. We achieved a best fit
with a model atmosphere and synthetic spectrum at $\Te=6440$~K,
$\logg=7.99$, $\nh=-3$, and $\log\ ({\rm Ca/He})=-8.1$, with other
heavy elements in solar abundances relative to the calcium abundance
(except for Na and Cr whose abundances had to be reduced by factors of
2 and 3, respectively). The predicted emergent spectrum provided an
excellent match to the whole energy distribution of G165-7, as well as
to most spectral features. By using the approximation of the linear
Zeeman effect in our synthetic spectrum calculations, we were able to
reproduce the splitting observed in several metallic lines. We also
found that, contrary to the original analysis of G165-7 by
\citet{WL1980}, hydrogen is present with an abundance of $
\nh\sim-3$. With an effective temperature $\sim1000$~K cooler than
previously found, the molecular formation of MgH can no longer be
completely ruled out. We have shown that the so-called ``asymmetric Mg
I profile'' might possibly be explained by a MgH molecular feature
instead of a quasi-static profile, although we cannot be entirely
certain of this interpretation until a coherent theoretical framework
becomes available. Other spectra of DZ stars with strong metallic
absorption features have been found in the SDSS \citep{harris03}.
Analysis of them may lead to more insight into the properties of
G165-7 and whether they are repeated in other DZ stars.

\acknowledgements{We are grateful to A.~Gianninas for a 
careful reading of this manuscript.  The authors thank P. Smith for
assistance with the spectropolarimetric observations.  Research into
magnetic stars and stellar systems at Steward Observatory is supported
by NSF grant AST 03-06080 to GDS.  This work was supported in part by
the NSERC Canada and by the FQRNT (Qu\'ebec). P. Bergeron is a
Cottrell Scholar of Research Corporation.

Funding for the SDSS and SDSS-II has been provided by the Alfred P.
Sloan Foundation, the Participating Institutions, the National Science
Foundation, the U.S. Department of Energy, the National Aeronautics
and Space Administration, the Japanese Monbukagakusho, the Max Planck
Society, and the Higher Education Funding Council for England.

The SDSS is managed by the Astrophysical Research Consortium for the
Participating Institutions. The Participating Institutions are the
American Museum of Natural History, Astrophysical Institute Potsdam,
University of Basel, Cambridge University, Case Western Reserve
University, University of Chicago, Drexel University, Fermilab, the
Institute for Advanced Study, the Japan Participation Group, Johns
Hopkins University, the Joint Institute for Nuclear Astrophysics, the
Kavli Institute for Particle Astrophysics and Cosmology, the Korean
Scientist Group, the Chinese Academy of Sciences (LAMOST), Los Alamos
National Laboratory, the Max-Planck-Institute for Astronomy (MPIA),
the Max-Planck-Institute for Astrophysics (MPA), New Mexico State
University, Ohio State University, University of Pittsburgh,
University of Portsmouth, Princeton University, the United States
Naval Observatory, and the University of Washington.
}

\clearpage

\input{tab1}

\input{tab2}

\clearpage

\figcaption[f1] {Comparison of absolute fluxes from two independent 
spectroscopic observations of G165-7 in the optical. The thick line is
the SDSS spectrum with a resolution of $\sim 3$ \AA\ FWHM, while the
thin line corresponds to the BLR spectrum at a slightly lower
resolution of $\sim 6$ \AA\ FWHM. The latter suffers from flux
calibration problems longward of 6500 \AA (also notice the presence of
the Earth's atmospheric A and B absorption bands in the BLR
spectrum). The most important features are labeled at the the top. The
presence of Zeeman splitting is clearly visible in the SDSS
spectrum. \label{fg:f1}}

\figcaption[f2] {Our best fit to the SDSS optical spectrum
with a non-magnetic model at $\Te=6440$~K, $\logg=7.99$,
and $\log\ ({\rm Ca/He})= -8.1$. All heavy elements are assumed
to be solar with respect to the calcium abundance, with the exception
of Na, Cr, and Fe whose abundances are reduced by factors of 2, 3, and
3, respectively. The hydrogen abundance is also set to zero.\label{fg:f2}}

\figcaption[f3] {Same as Fig.~\ref{fg:f2} but at a higher resolution. 
Note that the BLR spectrum is used here below 3800 \AA.\label{fg:f3}}

\figcaption[f4] {{\it Top panel:} Observed circular polarization ({\it thin
line}) and model polarization spectrum ({\it thick line}) for a mean
longitudinal field of 165~kG.  {\it Bottom panel:} Corresponding spectrum
obtained with a spectral resolution ($\Delta\lambda=17$ \AA) inadequate to
show the Zeeman splitting of the line cores.\label{fg:f4}}

\figcaption[f5] {Same as Fig.~\ref{fg:f3} but by taking into account
the presence of a weak magnetic field with a mean surface field
strength of $B_s\sim650$~kG in our model flux calculations. In this model,
the iron abundance is solar with respect to calcium. The insert in the
third panel from the top shows our prediction of a weak 
\halpha\ absorption feature when a small
abundance of hydrogen with $\nh=-3$ is included. \label{fg:f5}}

\figcaption[f6] {Our best fit to the energy distribution of G165-7
using models including metals as well as a weak magnetic field. The
observations are represented by the error bars while the corresponding
average model fluxes are shown by the filled circles ($BVRIJHK$, from
left to right) and by the open circles ($ugriz$, from left to right;
here the uncertainties have been set to 0.04 for clarity). The atmospheric
parameters are given in the figure.\label{fg:f6}}

\figcaption[f7] {Comparison of the observed and synthetic spectra in the
$5000-5300$ \AA\ region. Here, the MgH molecular opacity has been
included in our calculations ({\it solid line}), although an increase of the MgH
abundance by a factor of 20 was required to match the blue wing of
Mg~\textsc{i}. We also show our calculations without the inclusion of MgH 
({\it dashed line}).\label{fg:f7}}

\clearpage
\begin{figure}[p]
\plotone{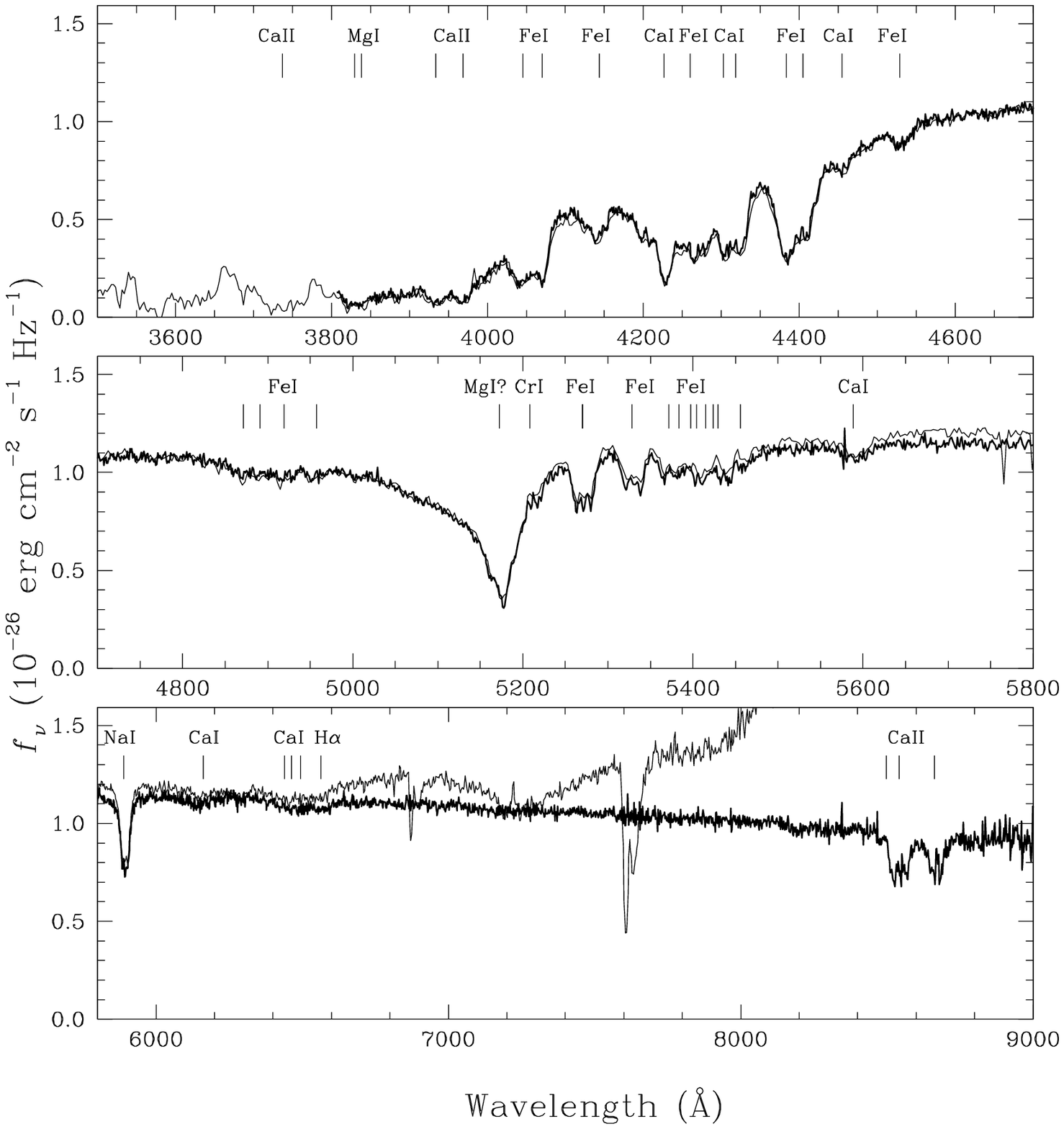}
\begin{flushright}
Figure \ref{fg:f1}
\end{flushright}
\end{figure}

\clearpage
\begin{figure}[p]
\plotone{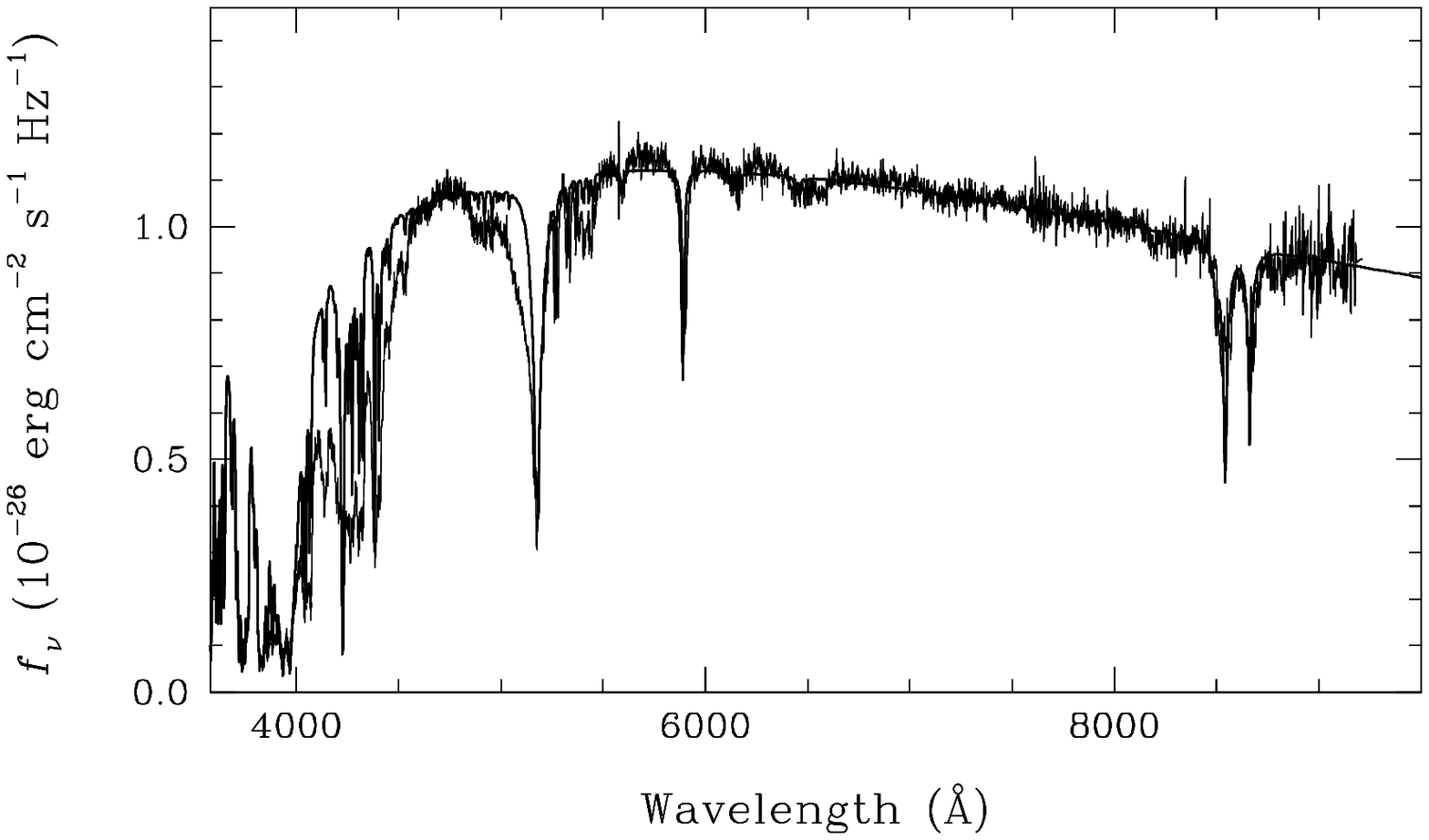}
\begin{flushright}
Figure \ref{fg:f2}
\end{flushright}
\end{figure}

\clearpage
\begin{figure}[p]
\plotone{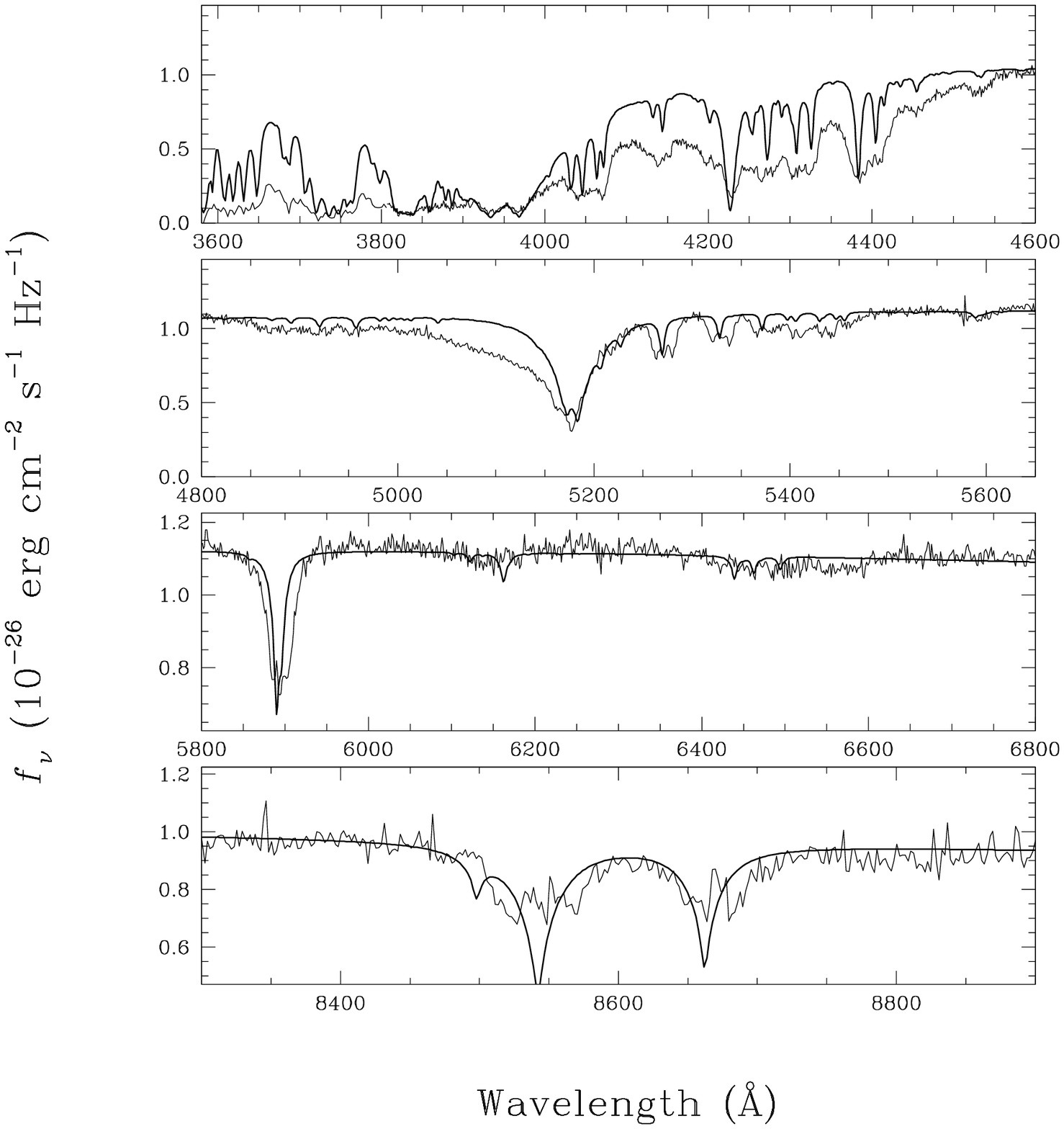}
\begin{flushright}
Figure \ref{fg:f3}
\end{flushright}
\end{figure}

\clearpage
\begin{figure}[p]
\begin{center} 
\includegraphics[angle=-90,width=6.0in]{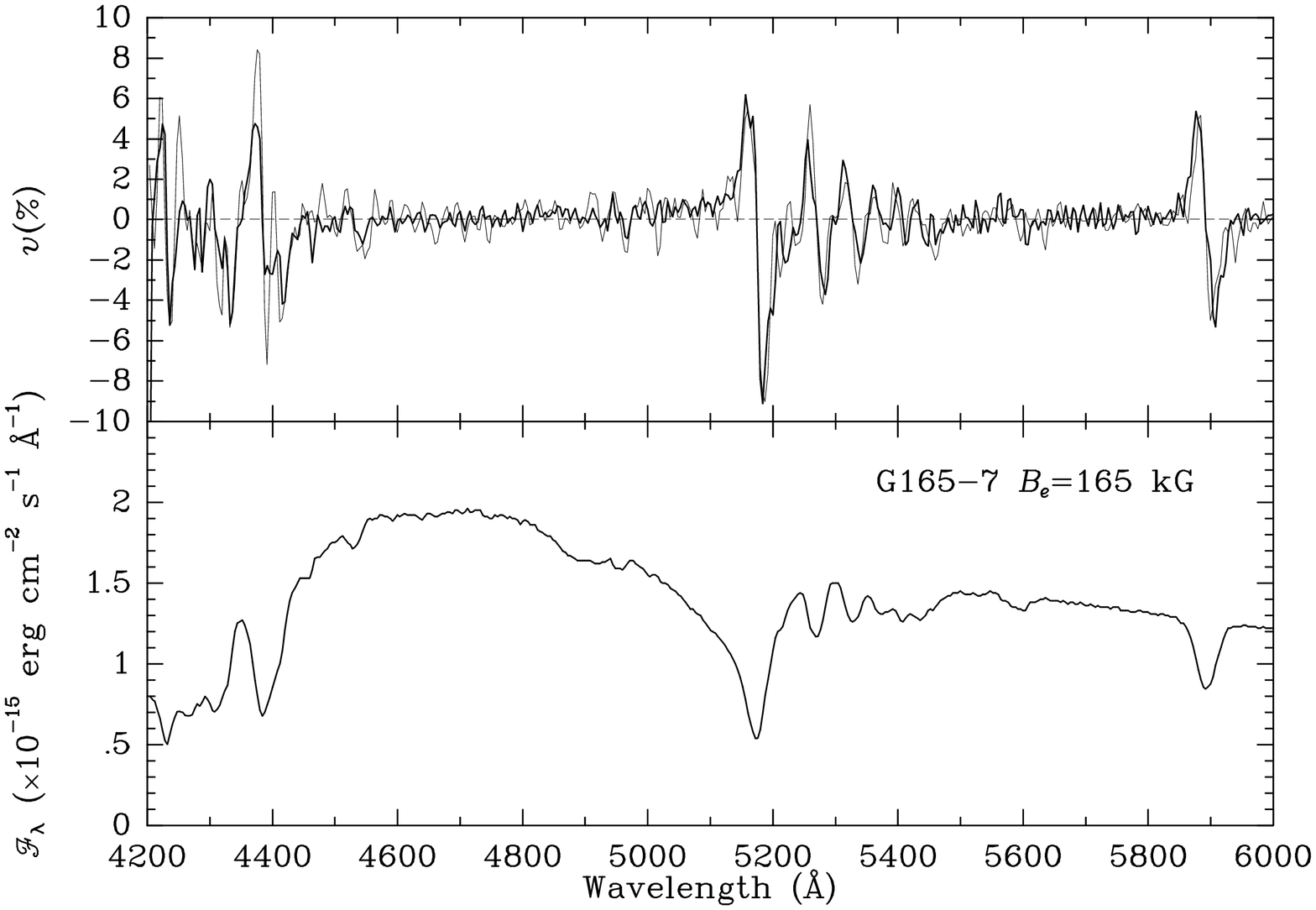} 
\end{center}
\begin{flushright}
Figure \ref{fg:f4}
\end{flushright}
\end{figure}

\clearpage
\begin{figure}[p]
\plotone{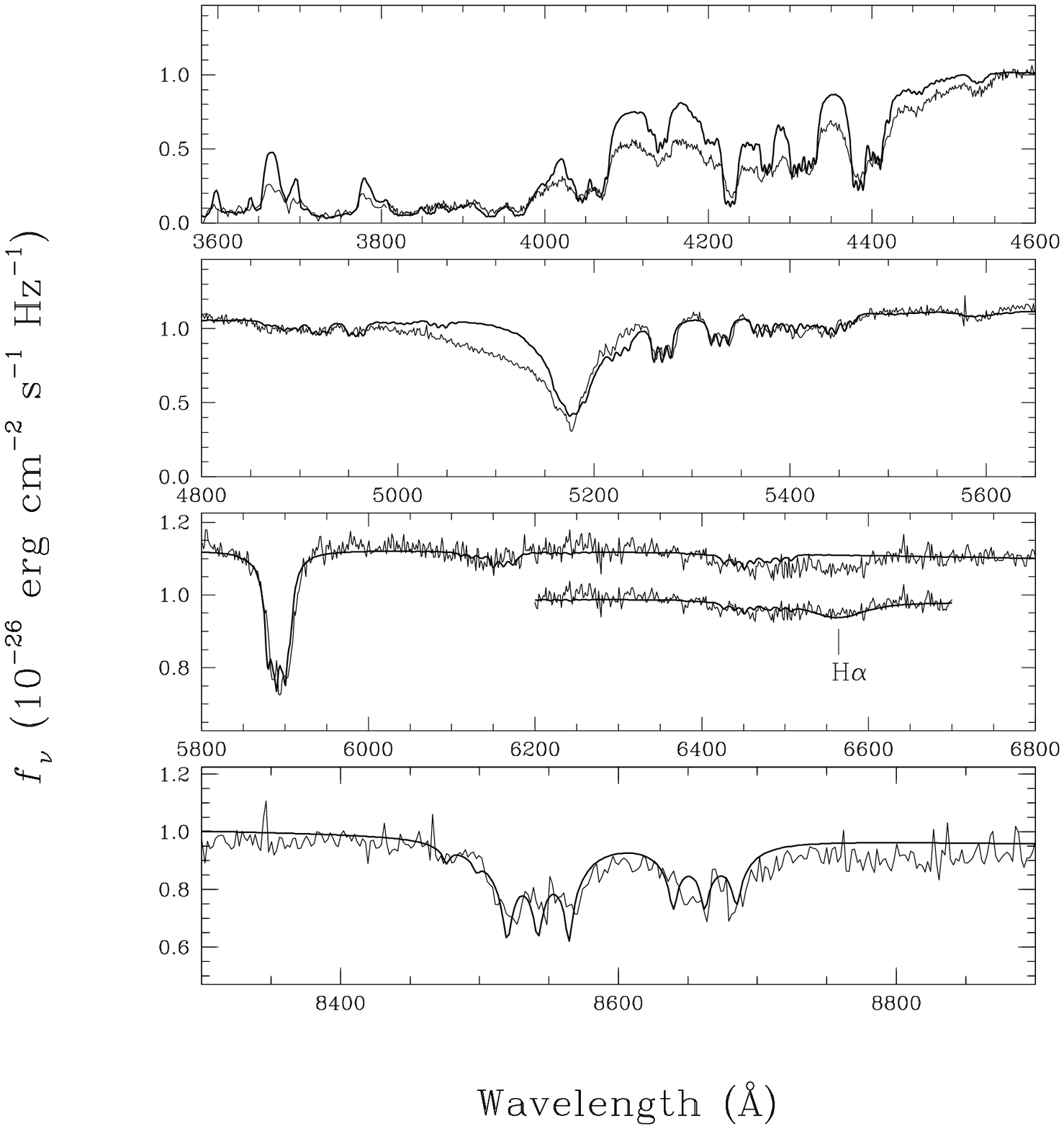}
\begin{flushright}
Figure \ref{fg:f5}
\end{flushright}
\end{figure}

\clearpage
\begin{figure}[p]
\plotone{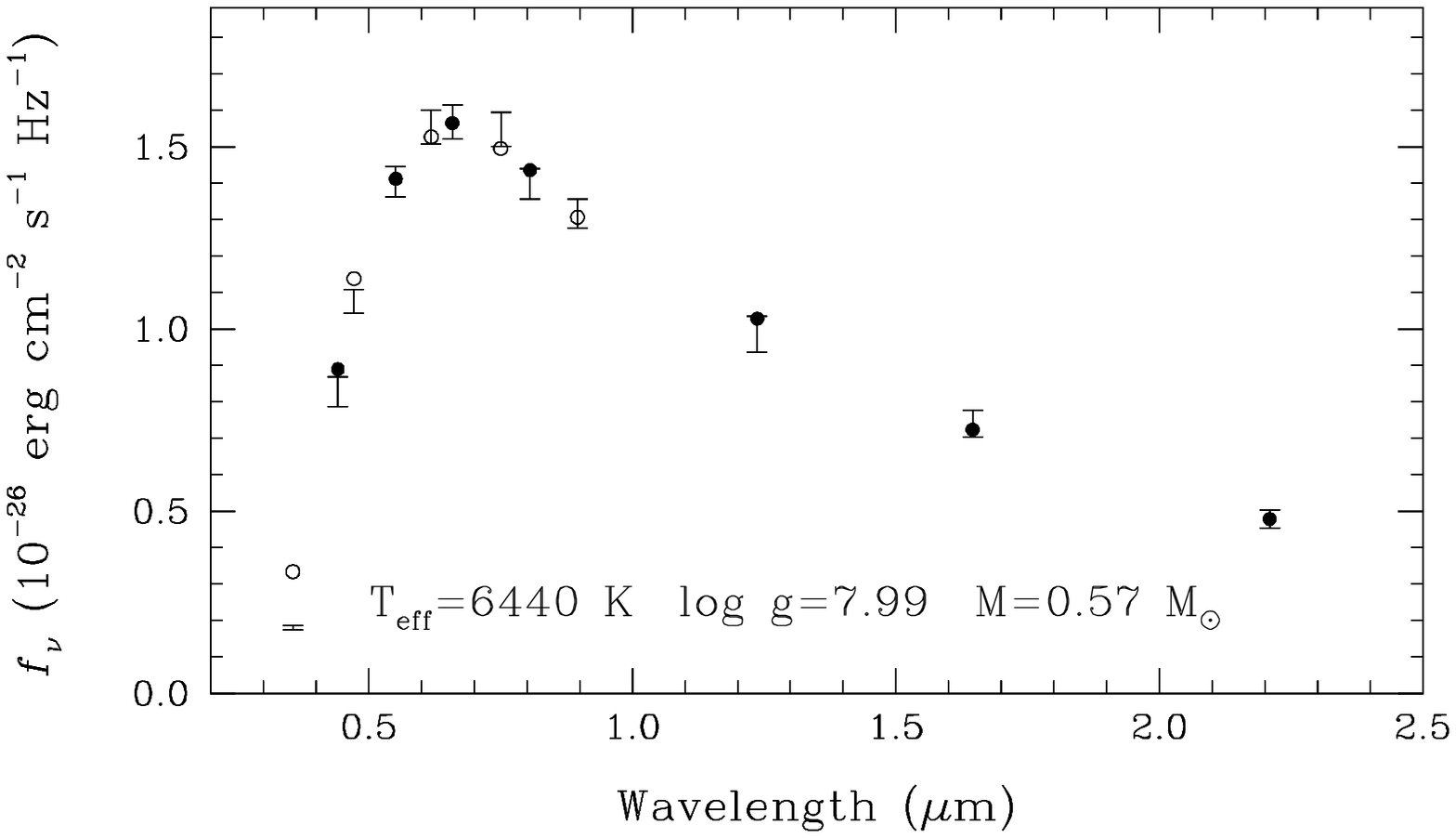}
\begin{flushright}
Figure \ref{fg:f6}
\end{flushright}
\end{figure}

\clearpage
\begin{figure}[p]
\plotone{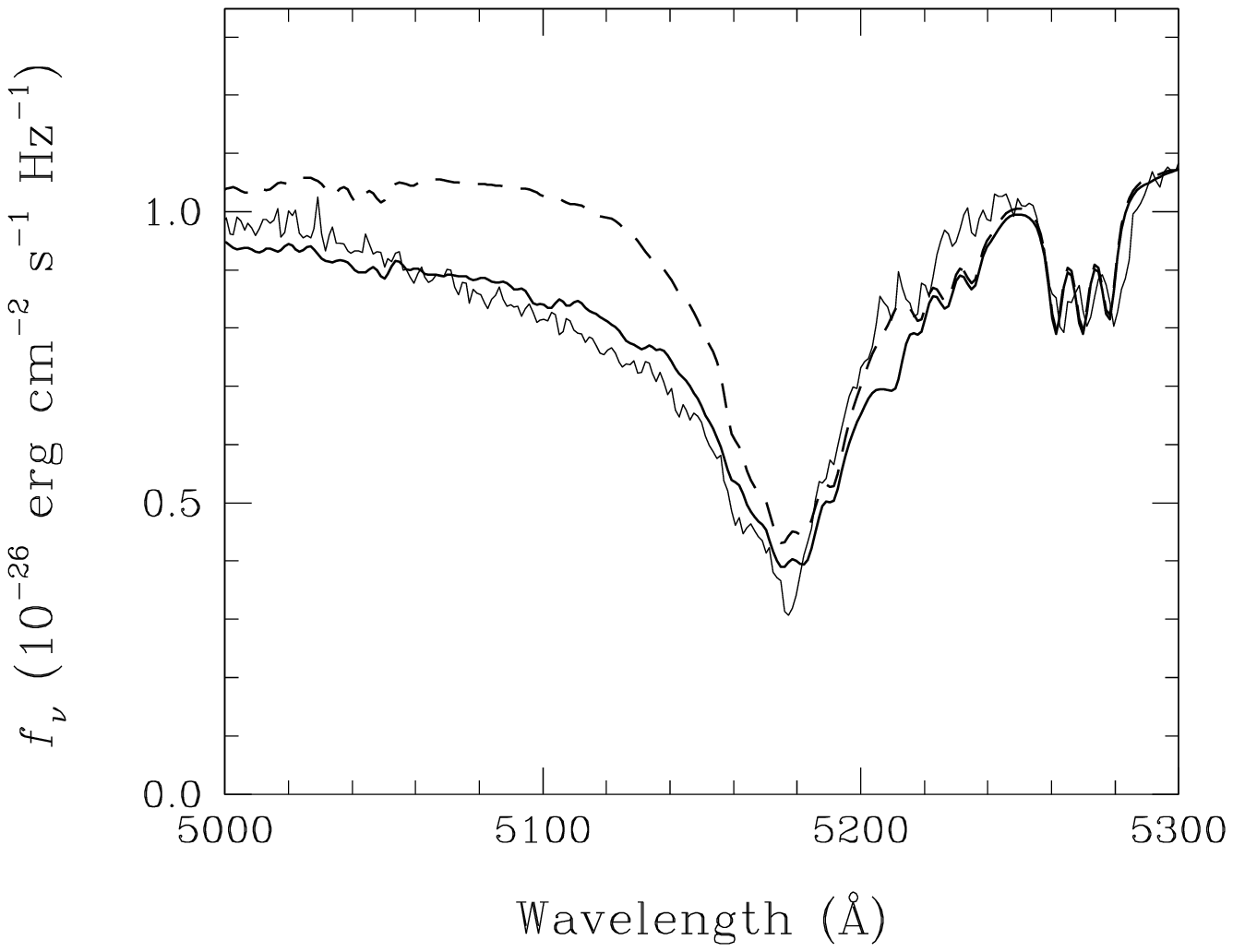}
\begin{flushright}
Figure \ref{fg:f7}
\end{flushright}
\end{figure}

\end{document}

%% file: tab1.tex
\clearpage
\begin{deluxetable}{rrrrrrrrrrrrc}
\tabletypesize{\footnotesize}
\tablecolumns{13}
\tablewidth{0pt}
\tablecaption{Observational data for G165$-$7\tablenotemark{a}}
\tablehead{
\colhead{$B$} & 
\colhead{$V$}& 
\colhead{$R$} & 
\colhead{$I$} &
\colhead{$J$} & 
\colhead{$H$} & 
\colhead{$K$} & 
\colhead{$u$} & 
\colhead{$g$} & 
\colhead{$r$} &
\colhead{$i$} & 
\colhead{$z$} & 
\colhead{$\pi$ (mas)}}
\startdata
16.73 & 16.03 & 15.74 & 15.60 & 15.50 & 15.36 & 15.34 & 18.27 & 16.32 & 15.92 & 15.93 & 16.10 & 33.4 \\
0.05 & 0.03 & 0.03 & 0.03 & 0.05 & 0.05 & 0.05 & 0.016 & 0.003 & 0.003 & 0.004 & 0.007 & 5.3 \\
\enddata
\tablenotetext{a}{Uncertainties are on the second line.
Cousins $BVRI$ and CIT $JHK$ photometry is from BLR, while the $ugriz$ photometry comes from SDSS.}

\clearpage
\end{deluxetable}

%% file: tab2.tex
\clearpage
\begin{deluxetable}{lccccc}
\tabletypesize{\scriptsize}
\tablecolumns{11}
\tablewidth{0pt}
\tablecaption{Atmospheric Parameters for G165$-$7}
\tablehead{Parameter & Value}
\startdata
$T_{\rm eff}$(K) & 6440 (210)\\
$\log g$         & 7.99 (0.29)\\
$M/M_{\odot}$    & 0.57 (0.17)\\
$\log$ H/He      & $\sim$$-3$ \\
$\log$ Na/He     &  $-$8.43 (0.15)\\
$\log$ Mg/He     &  $-$6.88 (0.15)\\
$\log$ Ca/He     &  $-$8.10 (0.15)\\
$\log$ Cr/He     & $<-$9.30 \\
$\log$ Fe/He     & $-$6.96  (0.15)\\
$B_s$(kG)        & $\sim$650\\
$B_e$(kG)        & $\sim$165\\
\enddata
\end{deluxetable}